\documentclass{elsarticle}

\usepackage{hyperref}

\usepackage[utf8]{inputenc}
\usepackage[T1]{fontenc}
\usepackage{amsthm,amsmath}
\usepackage{adjustbox}
\usepackage{booktabs}
\RequirePackage{hyperref}
\usepackage{multirow}
\usepackage[english]{babel}
\usepackage{xcolor}

\addto\captionsenglish{}

\journal{Physica A: Statistical Mechanics and its Applications}

\begin{document}

\begin{frontmatter}

\title{Modelling of temporal fluctuation scaling in online news network with independent cascade model}

\author[$a$]{Jan Cho{\l}oniewski}
\ead{choloniewski@if.pw.edu.pl }
\author[$a$]{Julian Sienkiewicz}
\ead{julas@if.pw.edu.pl}
\author[$b$]{Gregor Leban}
\ead{gregor.leban@ijs.si}
\author[$a$,$c$]{Janusz A. Ho{\l}yst\corref{*}}
\ead{jholyst@if.pw.edu.pl}
\address[$a$]{Center of Excellence for Complex Systems Research, Faculty of Physics, Warsaw University of Technology, Koszykowa 75, 00-662, Warsaw, Poland}
\address[$b$]{Artificial Intelligence Laboratory, Jo\v{z}ef Stefan Institute, Jamova 39, 1000, Ljubljana, Slovenia}
\address[$c$]{ITMO University, 49 Kronverkskiy av., 197101, Saint Petersburg, Russia}
\cortext[*]{Corresponding author.}

\begin{abstract}
We show that activity of online news outlets follows a temporal fluctuation scaling law and we recover this feature using an independent cascade model augmented with a varying \textit{hype} parameter representing a viral potential of an original article.
We use the Event Registry platform to track activity of over 10,000 news outlets in 11 different topics in the course of the year 2016. Analyzing over 22,000,000 articles, we found that fluctuation scaling exponents $\alpha$ depend on time window size $\Delta$ in a characteristic way for all the considered topics -- news outlets activities are partially synchronized for $\Delta>15\mathrm{min}$ with a cross-over for $\Delta=1\mathrm{day}$. 
The proposed model was run on several synthetic network models as well as  on a network extracted from the real data. Our approach discards timestamps as not fully reliable observables and focuses on co-occurrences of publishers in cascades of similarly phrased news items. We make use of the Event Registry news clustering feature to find correlations between content published by news outlets in order to uncover common information propagation paths in published articles and to estimate   weights of edges in the independent cascade model. 
While the independent cascade model follows the fluctuation scaling law with a trivial exponent $\alpha=0.5$, we argue that besides the topology of the underlying cooperation network  a  temporal clustering of articles with similar hypes is necessary to qualitatively reproduce the fluctuation scaling observed in the data.
\end{abstract}

\begin{keyword}
\texttt{Fluctuation scaling} \sep \texttt{Complex systems} \sep \texttt{Complex networks} \sep \texttt{Online media} \sep \texttt{Agent-based modelling} \\
Declarations of interest: none
\end{keyword}
\end{frontmatter}

\section{Introduction}
Rapid digitalization of our everyday life created possibilities to analyze previously inaccessible areas. Specifically, the emergence of Web 2.0 and social media (e.g., Twitter, blogosphere, Facebook) gave foundation to computational social sciences \cite{Conte2012}. A physicists' involvement can be seen as a successful one~\cite{Castellano2009, Kwapien2012, Helbing2014, DOrsogna2015, Wang2016, Perc2017}; examples of the latest studies leveraging the recent advancements are quantifying and modelling emotions~\cite{Chmiel2011,Garas2012,Cyberemo2017} and opinions~\cite{Sobkowicz2012} in online communities, network of scientific collaborations~\cite{Kuhn2014, Tomasello2017, Patania2017, Sienkiewicz2018}, dynamics of languages~\cite{Petersen2012}, or information diffusion and processing~\cite{Gomez-Rodriguez2012,Gomez-Rodriguez2013,breakingnews2016, He2017, Jalili2017}.

The social media revolution caused major changes in media industry. The Internet turned out to be an extremely effective medium for news stories changing the modern journalism -- publishing and spreading are more dynamic now, news is produced and updated continuously; also the gatekeeping function of journalists has been reduced since every person or entity can directly post messages on social media. For data scientists, the shift towards digitalized environment created an opportunity to quantitatively analyze activity and content of news outlets. One can perceive online news ecosystem as a complex network consisting of news outlets producing information themselves or mimicking/processing information observed in other sources (such as news agencies, social media, or other news outlets). Such  dynamical systems often follow some kind of a fluctuation scaling law~\cite{fronczak} which can be identified to provide conclusions about degree of units' temporal~\cite{Choloniewski2016} or spatial~\cite{Petri2013} synchronization in different scales.

The study presented here is an analysis of 22 million articles with one of 11 keywords published in 2016 gathered by \textit{EventRegistry.org}~\cite{er}, a global media monitor. The main aim of our study is to report the temporal fluctuation scaling found in the dataset and check whether the statistical property of media activity can be described with an epidemic-like process. The observed type of the fluctuation scaling can give hints on underlying system dynamics. We show that a possible explanation of the observed fluctuation scaling could be an underlying independent cascade model taking place on a complex network. We track propagation of news stories to approximate a topology of the network. While in many situations information is propagated as a straightforward connection, like retweets, in other cases it may mutate, change its form or sentiment, or even become a mix of information from different sources. In journalism, explicitly mentioning a cited source is considered a good practice but, due to the competitiveness of the industry, it is not always met. Without reliable information about which news item was published first (as it might be a matter of seconds), whether given piece of content was produced or copied, or even whether its original source is being observed, we decided to apply natural language processing methods to meaningfully group published news items across various news outlets~\cite{Jacobi2016, jch}. Aggregating results across an another dataset (over 14,000 events observed between 1.5.2017 and 8.5.2017 limiting to articles written in English) uncovered a content correlation network which was used to simulate the process of news spreading. We show that the independent cascade model in its common form does not provide full explanation of the fluctuation scaling observed in the data, and postulate a specific  news item feature -- \textit{hype} -- which represents its intrinsic viral potential and causes the model to indicate realistic scaling exponents.

There is an abundance of recent methods to uncover network basing on the independent cascade model (typically continuous time~\cite{Barbieri2013, PougetAbadie2015, Yu2017}) which assume exactly measured timestamps; here we utilize a discrete approach and focus mostly on cascade sizes and publishers co-occurrences. Also, modelling of fluctuation scaling observed in online social communities is lastly a vivid topic of research (mostly with a random diffusion model~\cite{Sano2015}, also with a time varying scale parameter to model a word occurrences fluctuations scaling in blogosphere~\cite{pretaylor2016}). Moreover,~\cite{Alessandretti2017} and~\cite{Pozzana2017} already cover the topic of varying attractiveness of a message propagating in a complex network, however it is attributed to its producer not the message itself. To the best of our knowledge no other study addresses the temporal fluctuation in the news outlet network nor in the independent cascade model.

The rest of the document is structured as follows: first, we describe the dataset in detail and gives its basic statistics (Section~\ref{sec:data}); then we show that the media outlets activity follows a fluctuation scaling law, and report how scaling exponents depend on timescale and unit size (Section~\ref{sec:tfs}); next, we describe a reconstructed network of news outlets based on content correlations and basic features of such a system (Section~\ref{sec:net}); we finish the results section with showing how the independent cascade model can recover stylized facts observed in the data (Section~\ref{sec:model}). Finally, we draw conclusions, discuss, and summarize the work (Section~\ref{sec:discussion}). In three appendices, we consider basic statistical properties of investigated datasets (\ref{app:fits}), and introduce algorithms applied to compute fluctuation scaling exponents (\ref{app:tfs}) and extract publishers network (\ref{app:network}).

\section{Data and its basic statistics}
\label{sec:data}
\begin{figure}
\centering
\includegraphics[width=\textwidth]{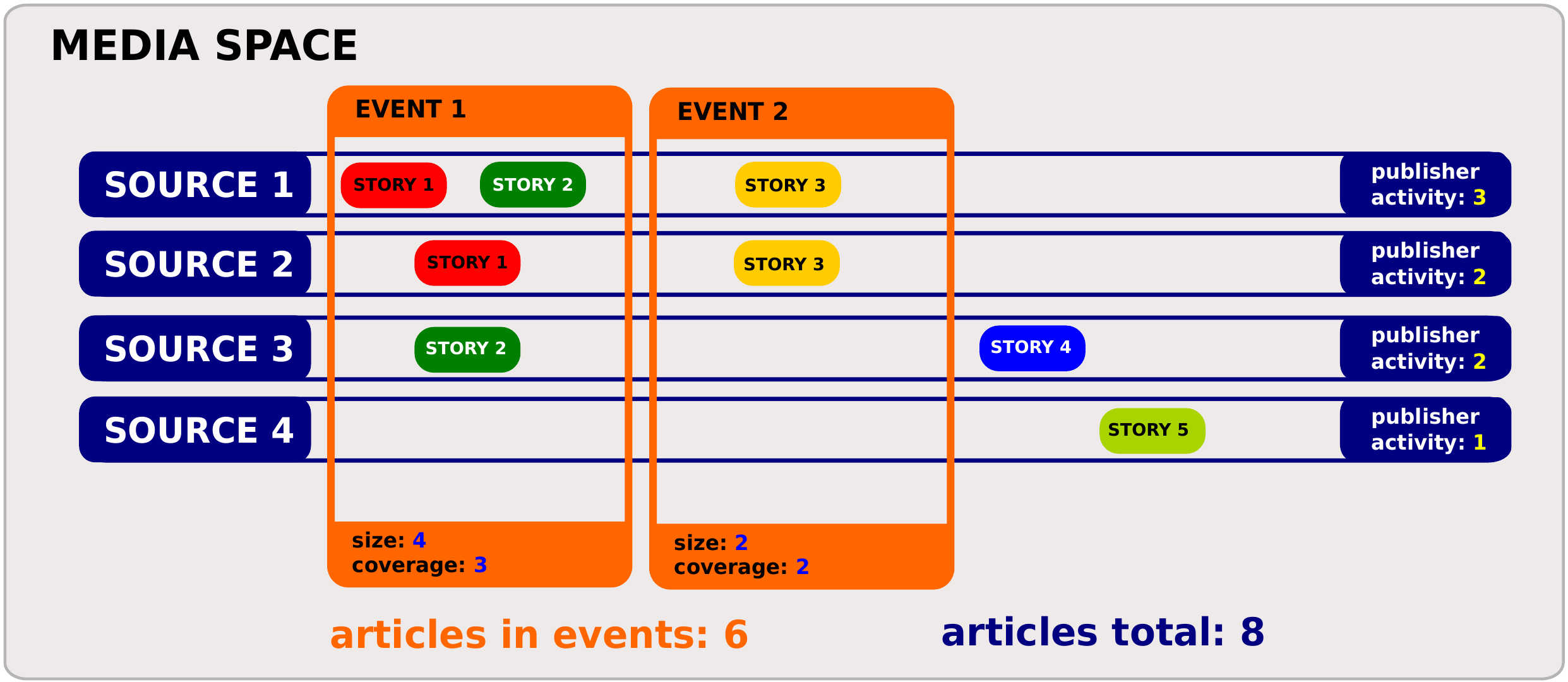}
\caption{A visualization of the dataset structure and its considered features -- event size (number of articles assigned to a given event), publisher activity (number of articles published by a given outlet), and event coverage (number of publishers having at least one article assigned to a given event). Story $i$ is an article from an $i$-th cascade (see Methods) of very similar articles published by different sources. \textit{Source 4} is an example of a publisher with no articles assigned to an event; other Sources are publishers in events. The Event Registry platform tracks only events consisting of at least 5 articles.}
\label{fig:arts_pubs_events_example}
\end{figure}

Event Registry platform monitors RSS feeds of over 25,000 news outlets for articles in 35 major languages from around the world, forms temporal clusters of similar articles (\textit{events}), and extracts metadata about each event (such as involved entities, recognized Wikipedia-based concepts, location) ~\cite{er}. A list of covered sources includes the biggest players in the news industry like online versions of tabloids (e.g. \textit{dailymail.co.uk}) or national daily newspapers (e.g. \textit{welt.de}), international news aggregators (e.g. \textit{www.msn.com}), as well as numerous local or otherwise narrowly-focused news providers. A cluster of at least 5 similar articles, published within an interval between the newest and the oldest article of at most 4 days, forms an \textit{event}. The dataset for the analysis consists of around 1,500,000 events (over 22,000,000 articles) published in 2016 and mentioning one of 11 subjectively chosen entities and concepts. We selected mostly keywords which were connected to major events of the year 2016. The first group were three major figures in the presidential campaign in the United States of America (the previous president \textit{Barack Obama}, the Democratic Party's presidential candidate -- \textit{Hillary Clinton}, and the Republican Party's candidate -- \textit{Donald Trump}); the second group consisted of parties most actively involved in Brexit talks (\textit{European Union}, \textit{United Kingdom}, \textit{Germany}, \textit{France}). The \textit{democracy} keyword alludes to both of the groups as those revolved around the voting process. \textit{Association football} was selected because of its stable popularity in Europe and a major event in 2016 (UEFA European Championship). \textit{Poland} and \textit{Argentina} were selected as a kind of a baseline -- both are countries not very visible in the international media, and devoid of globally impactful happenings in the year.
In this Section we consider distributions of \textit{publisher activities} (article number by publisher), and \textit{event sizes} (article number by event) and \textit{event coverages} (distinct publisher number by event). For a visual explanation -- see Fig~\ref{fig:arts_pubs_events_example}.

In Table~\ref{tab:basic}, we show the number of articles and publishers (both total and assigned to events only) and number of articles assigned to an event for each concept. All columns in the table are strongly correlated with each other (\textit{r}-Pearson coefficients $>.9$ with \textit{p}-values $<10^{-4}$), and the total number of articles is the best proxy to describe popularity of the keyword (\textit{r}$>.929$ with all the other columns) among the columns. The mean fraction of articles assigned to events is $(43\pm 2)\%$, and the mean percent of publishers with at least one event-related article is $(79\pm 5)\%$ (both errors as a standard deviation in a given sample). Percents of articles in events and publishers in events have a relatively low sample-standard-deviation-to-mean ratio (below $0.06$) comparing to other columns (above $0.12$); it is not surprising, as the two mentioned values are normalized (thus intensive, opposing to the rest of the columns which contain extensive values). It also proves that, while a popularity varies among keywords, each keyword is covered in a similar way. 
Distributions of the aforementioned values were all fat-tailed but we were unable to determine what function is the best description. For further considerations of the distributions fitting -- see~\ref{app:fits}.

\begin{table}
\centering
\begin{adjustbox}{center}
\begin{tabular}{l|rr|rr|r}
\toprule
{concept} &     Articles &  in events &  Publishers &  in events &  events \\
\midrule
Barack Obama         &  1,475,957 &          682,169 (46.2\%)&  12,067 &            9,941 (82.4\%) &   83,061 \\
Hillary Clinton      &  1,302,358 &          586,636 (45.0\%) &  10,311 &            8,166 (79,2\%) &   54,095 \\
Donald Trump         &  2,100,420 &          932,706 (44.4\%) &  11,700 &            9,506 (81.2\%)&   84,575 \\
European Union       &  2,112,576 &          942,675 (44.6\%) &  11,989 &            9,740 (81.2\%) &  137,272 \\
United Kingdom       &  3,432,419 &         1,497,603 (43.6\%)&  15,488 &           12,966 (83.7\%) &  265,734 \\
Germany              &  3,706,909 &         1,546,498 (41.7\%) &  15,143 &           12,274 (81.1\%) &  278,536 \\
France               &  3,247,340 &         1,397,352 (43.0\%) &  14,751 &           12,073 (81.8\%) &  227,404 \\
Poland               &   511,151 &          207,339 (40.5\%) &  10,112 &            7,219 (71.4\%)&   42,850 \\
Argentina            &  1,099,618 &          501,235 (45.6\%) &  10,357 &            7,788 (75.2\%) &   85,768 \\
democracy            &  1,048,966 &          431,787 (41.1\%) &  11,520 &            8,974 (77.9\%) &   90,742 \\
association football &  2,321,356 &          970,251 (41.7\%) &  13,926 &           10,987 (78.9\%) &  184,988 \\
\bottomrule
\end{tabular}
\end{adjustbox}
\caption{Basic properties of 11 examined concepts: numbers of articles, articles assigned to events, publishers, publishers which published at least one article assigned to an event, and events associated to each concept in the dataset. Percent values are calculated in relation to corresponding total numbers of articles or publishers.}
\label{tab:basic}
\end{table}

\section{Results}
\subsection{Temporal fluctuation scaling in the dataset}
\label{sec:tfs}
The fluctuation scaling law (often called \textit{Taylor's law} after the famous L. R. Taylor's paper~\cite{Taylor1961}) is an empirical law observed in complex systems which consist of differently sized, otherwise similar, units. The law binds means and standard deviations of units' activity (or, in general, a positive additive value characterizing each unit) in a form of a power law: $\mu\sim \sigma^\alpha$. The value of the exponent $\alpha$ is known to indicate a degree of units' synchronization in the observed system. It is common for $\alpha$ to be in range from 0.5 (uncorrelated units) to 1.0 (perfect synchronization, e.g., strong external force) but also higher values have been reported~\cite{fronczak} and theoretically approved~\cite{pretaylor2016}. Intermediate values are usually interpreted as a mixture of correlated and uncorrelated dynamics governing the system.
There are two variants of the law. The ensemble fluctuation scaling can be observed if one groups units by a scale (\textit{size-like}) parameter and consider means and variances calculated in each group. The temporal variant requires an observation of units' activity which can be spotted over time. In this variant, the observation period is divided into time windows of size $\Delta$, then the activity is aggregated in each time window for each unit. The statistics $\mu$ and $\sigma$ are calculated for each unit separately over all time window~\cite{Eisler2008}. In this study we focus on the latter variant and describe it in detail in ~\ref{app:tfs}. Particularly, it is interesting for us to look at units' activities aggregated in different time windows (i.e. at different timescales). 

\begin{figure}
\centering
    \includegraphics[width=\textwidth]{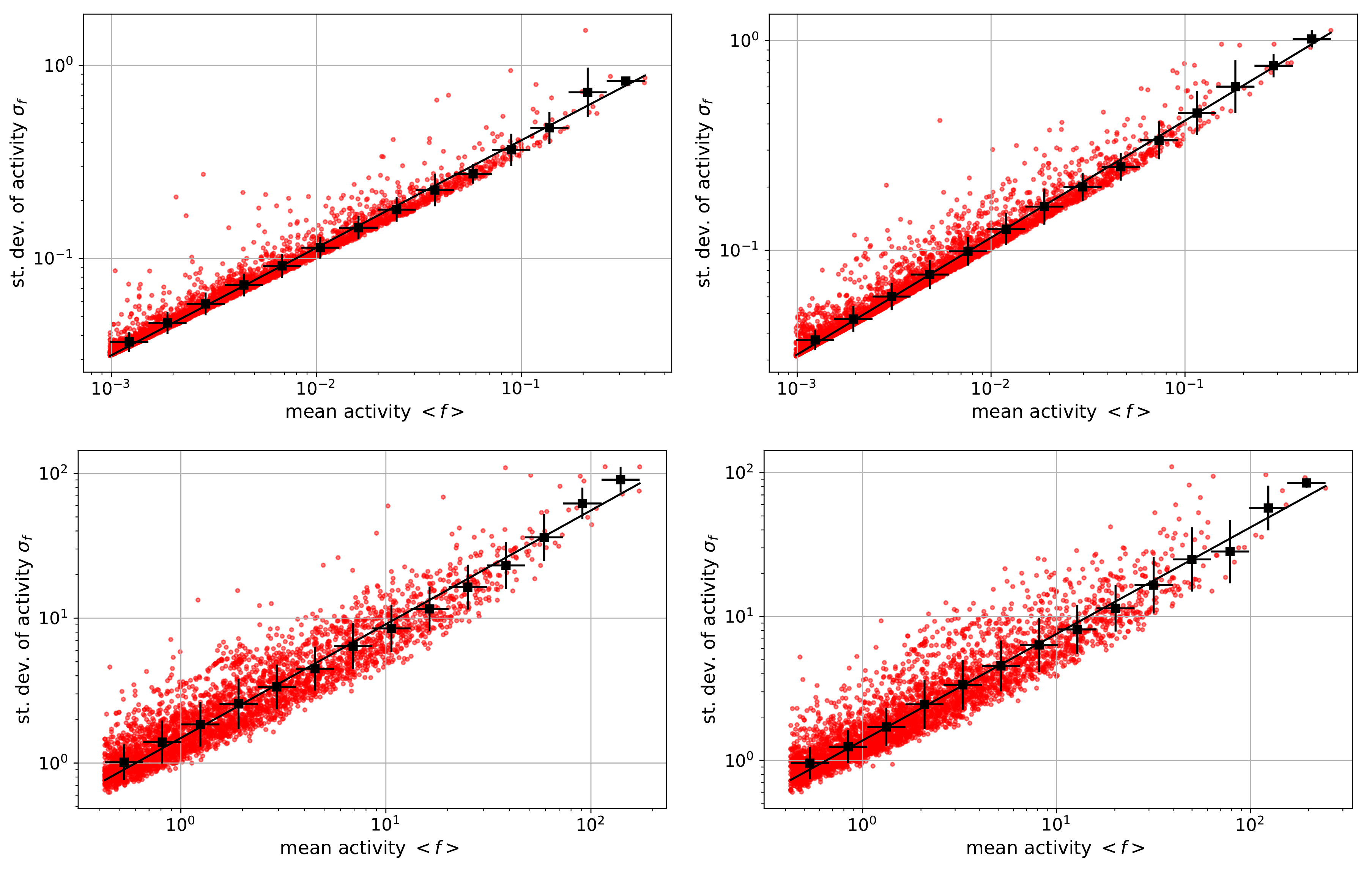}
\caption{Temporal fluctuation scaling law of news outlets activity for time windows size $\Delta\in \{ 10 \mathrm{min}, 3 \mathrm{days}\}$ for top and bottom, respectively. Articles about (left) \textit{European Union} and (right) \textit{association football}. X-axes -- an average number of articles mentioning a given concept per fixed time window $\Delta$ for a given publisher, Y-axes -- a standard deviation of a number of the articles. Only publishers with more than 1 article per week on average were considered ($N_{pubs}=3349$ and $N_{pubs}=3974$, respectively). Each red point represents one publisher, black points stand for a mean standard deviation in a given logarithmic bin, black line is a power law fit. Slopes are (left) $\alpha(10 \mathrm{min})=0.557\pm 0.022$, $\alpha(3 \mathrm{days})=0.786\pm 0.043$, (right) $\alpha(10 \mathrm{min})=0.557\pm 0.013$, $\alpha(3 \mathrm{days})=0.740\pm 0.037$.}
\label{fig:tfs_single}
\end{figure}

We observed the temporal fluctuation scaling for all analyzed keyword and for all sizes of time windows. Figure~\ref{fig:tfs_single} (left -- \textit{European Union}, right -- \textit{association football}) shows scatter plots of standard deviation $\sigma$ versus mean $\mu$ value of publisher activities (number of articles in a time window) for two selected time windows $\Delta\in \{ 10 \mathrm{min}, 3 \mathrm{days}\}$. The slopes of fitted lines are  fluctuation scaling exponents $\alpha(\Delta)$. In each of the selected time window sizes, exponents are nearly the same for both concepts. For the longer time window, points are more scattered around the fitting line and the scaling exponents $\alpha>0.7$ (for the shorter one -- $\alpha\approx0.55$).

\begin{figure}
\centering
    \includegraphics[width=\textwidth]{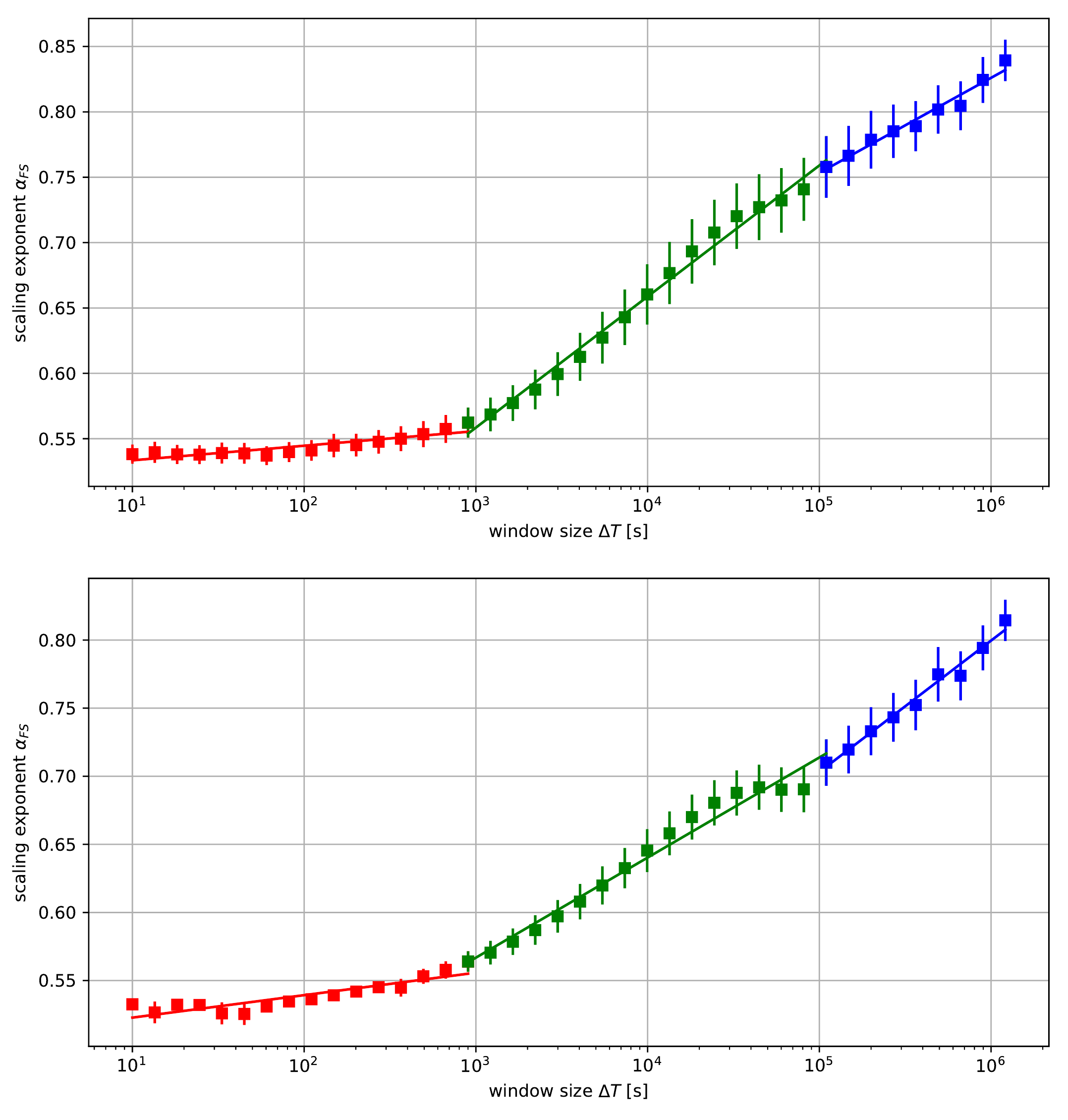}

\caption{The temporal fluctuation scaling exponent: $\alpha$ is nearly monotonically increasing with time window size $\Delta$; character of the dependence is similar for all the analyzed concepts. (top) \textit{European Union}, (bottom) \textit{association football}. X-axis is in log scale. Lines are fit of logarithm function $\alpha \sim \log\Delta$. Breakpoints are set manually to $\Delta=15$ min and $\Delta=1$ day. Slopes are (sequentially red, green, and blue line): (left) $\gamma_1=0.011\pm 0.002$, $\gamma_2=0.100\pm 0.003$, $\gamma_3=0.073\pm 0.006$, (right) $\gamma_1=0.016\pm 0.002$, $\gamma_2=0.073\pm 0.003$, $\gamma_3=0.096\pm 0.006$.}
\label{fig:tfs_dt}
\end{figure}

\begin{figure}
    \centering
    \includegraphics[width=\textwidth]{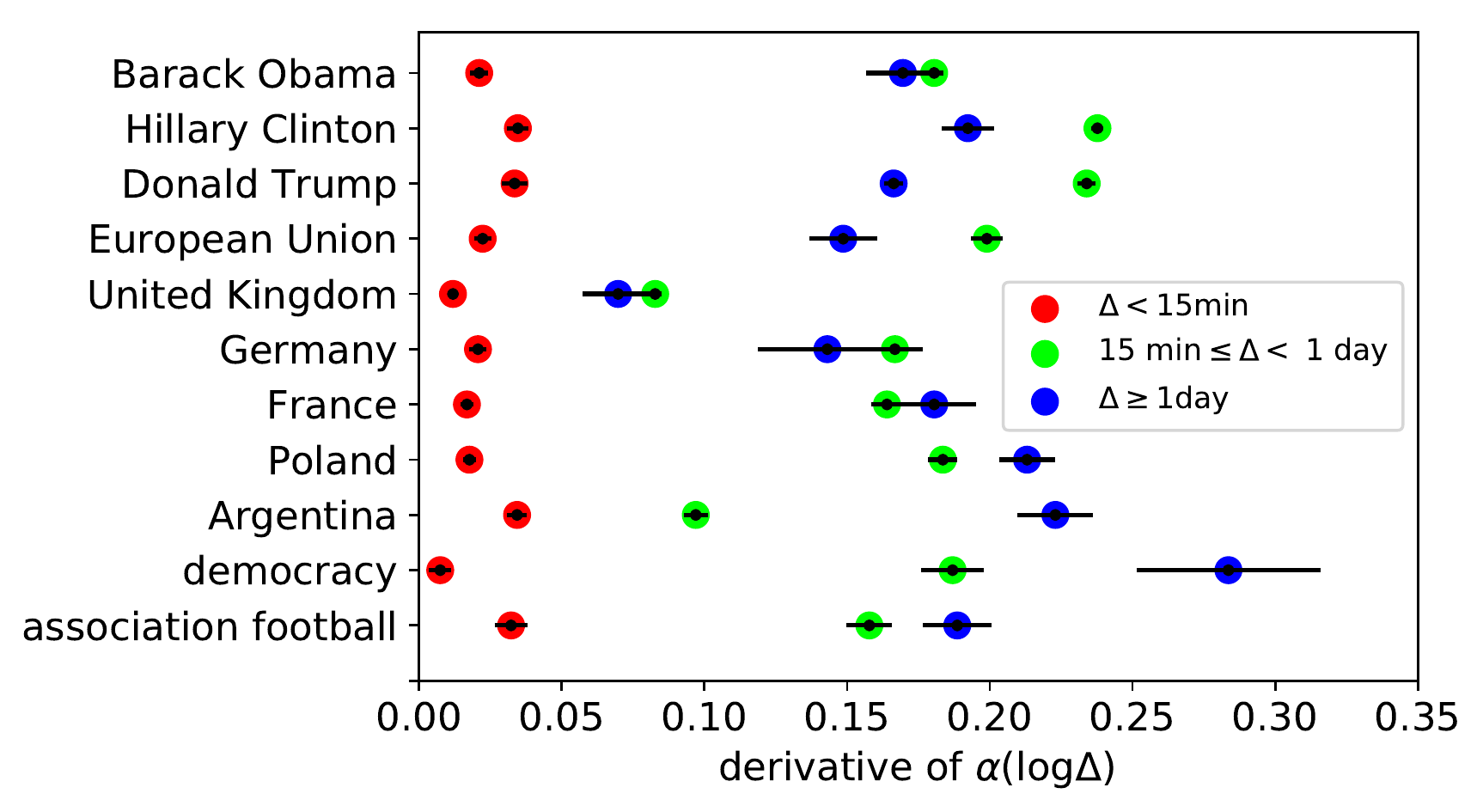}
\caption{Slopes of temporal fluctuation scaling $\alpha(\log\Delta)$ for different timescales and concepts. In shorter timescales ($\Delta<15\mathrm{min}$) the slopes are similar for all the analyzed concepts. For longer timescales, the slopes are varied and, in few cases, separated for $15\mathrm{min}\leq\Delta\leq1\mathrm{day}$ and $\Delta>1\mathrm{day}$. Errors as a standard deviation of the slope coefficient.}
\label{fig:tfs_fits}
\end{figure}

The dependence of $\alpha$ on $\Delta$ for the aforementioned keywords is shown in Fig.~\ref{fig:tfs_dt} presenting three linear regimes with different slopes. A piecewise linear fit was applied to recover slopes in the regimes. Automatically detected breakpoints varied slightly for different keywords thus, for the sake of comparisons, we manually set them to $15\mathrm{ min}$ and $1\mathrm{ day}$ as these values were the most common in the automatic breakpoint detection. The exponent $\alpha$ grows (nearly) monotonically with $\Delta$. For short time windows (up to $\sim15$ min), $\alpha$ is close to $0.5$ and growing with a slope $\gamma_{1}\approx0.01$ per decade. For longer time windows the growth is much faster ($\gamma_{2}\approx\gamma_{3}\approx0.09$). Regime slopes for all keywords in all regimes can be found in Fig.~\ref{fig:tfs_fits}.

\subsection{Extracted publishers network}
\label{sec:net}
In this section we will use Event Registry data to extract publishers network that will be further used in the next section to run a model of news cascades that reproduces the fluctuation scaling observed in the previous section. Nodes represent news outlets and weights of directed edges represent tie strengths. While not all news outlets can be observed and the exact propagation paths are impossible to follow, we hope to uncover meaningful connections between various publishers by processing reasonable amount of data related to contents of published articles. The full procedure of extracting the publishers network is described in ~\ref{app:network}.

The network will be an environment for simulations of the independent cascade model (see the next section) thus we need to keep it reasonable size to make simulations faster. The resulting graph has 5,719 nodes and 1,329,030 edges.
The vast majority of nodes was active only few times during the analyzed period, thus for readability and computational feasibility of simulations, we decided to prune the recovered network leaving only edges with $u_{ij}>0.5$ in the network recovered from the full dataset. The model was run on a giant component of the pruned graph (1,037 nodes and 4,150 edges).

\begin{figure}
\centering
    \includegraphics[width=\textwidth]{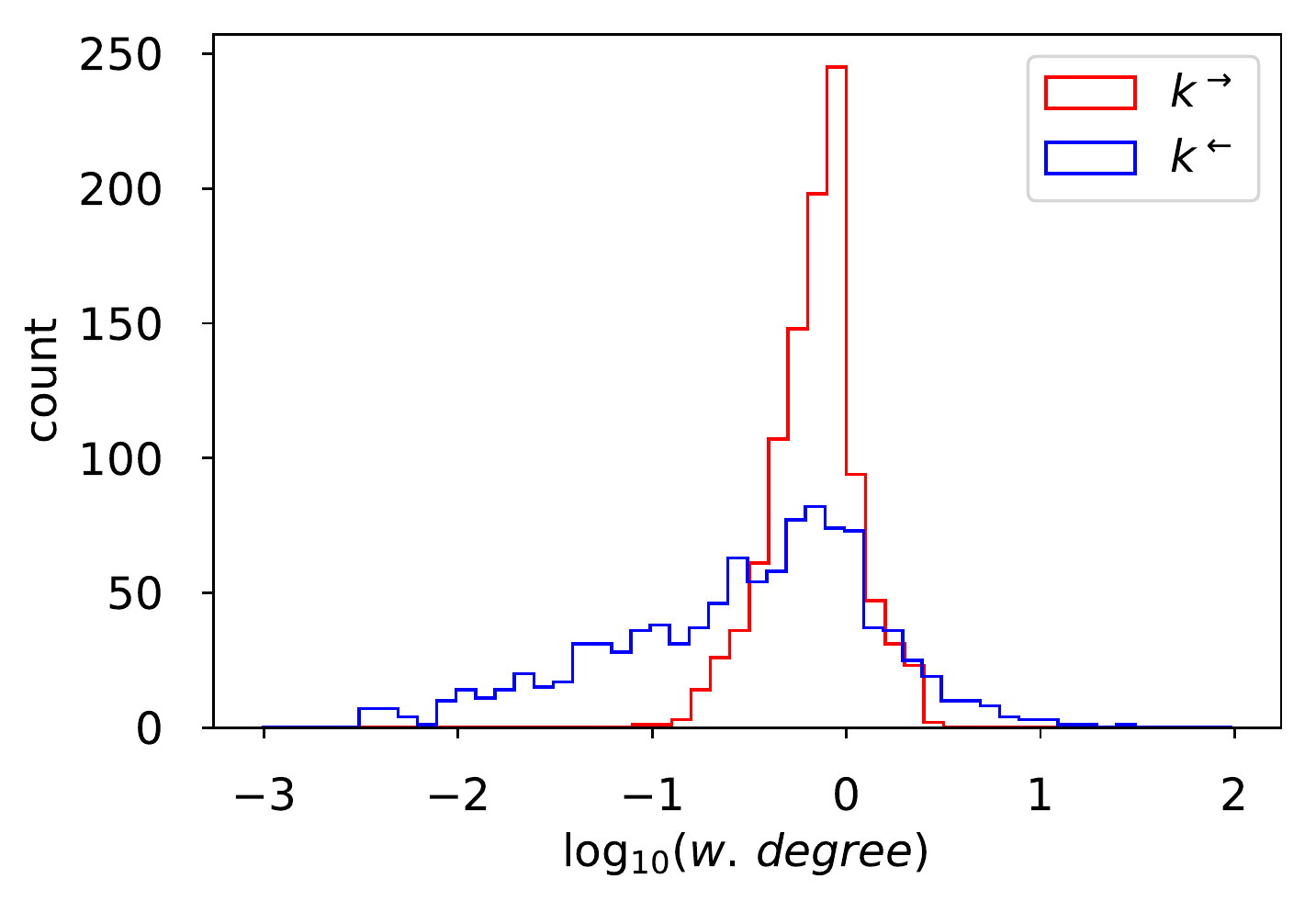}
\caption{The histograms of weighted degree of nodes in the pruned publishers network do not follow a power law; in-degrees distribution is wider than the out-degrees distribution. Weighted in- $k^{\leftarrow}$ and out-degrees $k^{\rightarrow}$ are defined in ~\ref{app:network})). Logarithmic bins.}
\label{fig:deg_hist}
\end{figure}

A logarithmic binning of weighted degree distribution of the pruned graph (Fig.~\ref{fig:deg_hist}) shows that the in-degree distribution is wider than the out-degree distribution. Degrees of nodes in real networks are often power-law distributed but in the recovered graph there are nodes with a relatively high number of neighbors but not as much lowly connected nodes as one would expect. This is probably caused by the filtering of publishers with low activity.

\begin{figure}
\centering
\begin{adjustbox}{center}
    \includegraphics[width=1.35\textwidth]{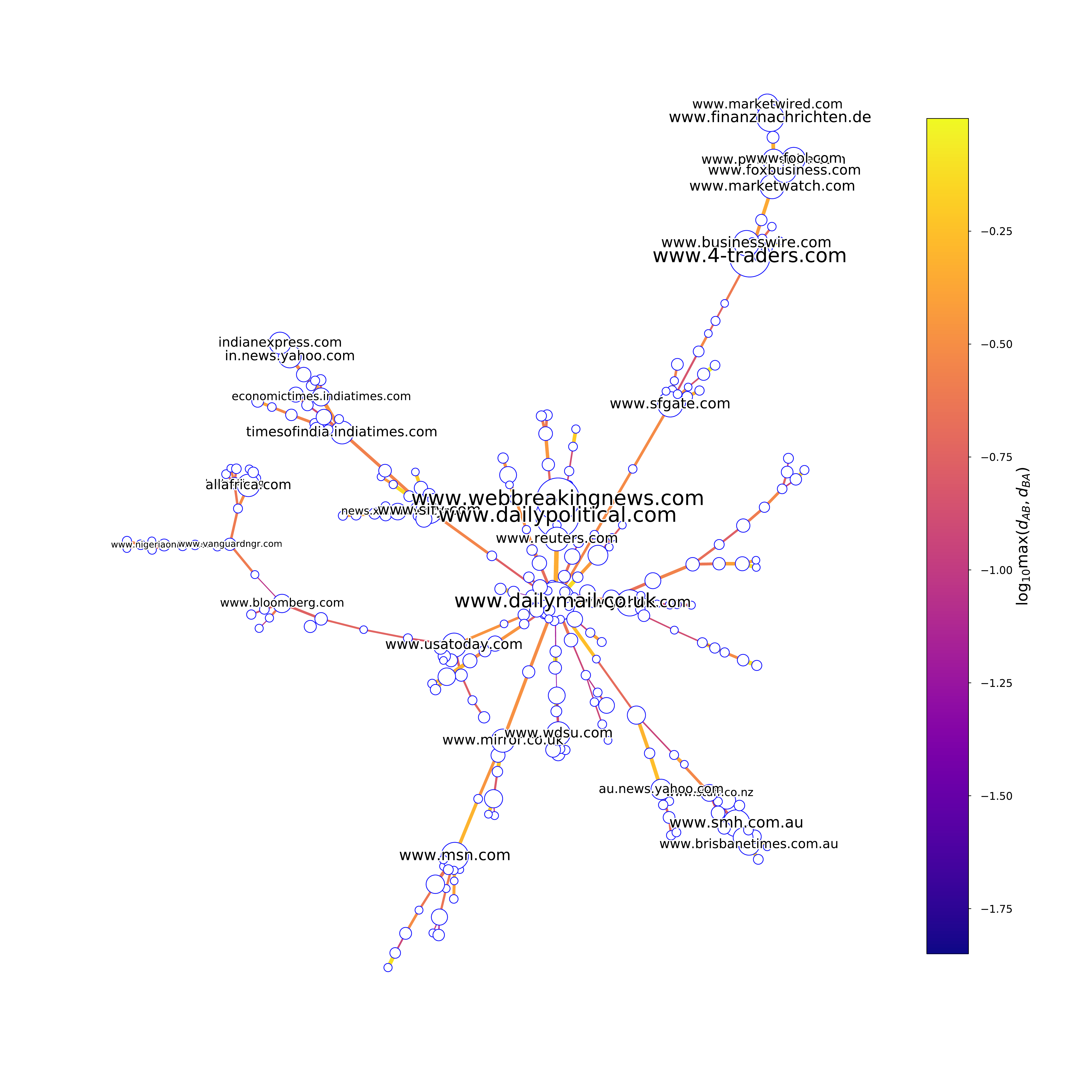}
\end{adjustbox}
\caption{Extracted publishers network: the network consists of geographical and topical clusters; local hubs and important information flow channels are visible. Maximum spanning tree of the giant component of the extracted publishers network based on $\max{(d_{ij},d_{ji})}$. Only publishers with $N_{i}>5$; only edges with $\max{(d_{ij}, d_{ji})}>0.01$, color of edge depends on logarithm of $\max{(d_{ij},d_{ji})}$. Size of a node is proportional to its $N_{i}$. Width of an edge is proportional to its $u_{ij}$ Labels were shown for nodes with a degree in the MST over 2. The graph has 289 nodes and 288 edges.}
\label{fig:net_vis}
\end{figure}

A maximum spanning tree of the giant connected component of the pruned network is presented in Fig.~\ref{fig:net_vis} (only nodes with $N_{i}>5$). The MST was calculated for an undirected graph with weights set to $\max{(d_{ij},d_{ji})}$. Geographical clustering of nodes for the major English-speaking countries (UK, USA, India, Australia, New Zealand) with addition of English versions of major local outlets (China, Africa) is clearly visible. While the visualization does not allow to determine direction of an edge, the most active connections reveal a few major information flow channels. 

\subsection{Model}
\label{sec:model}
The independent cascade model is an epidemic model run on a complex network. Nodes represents individuals (persons, social network users, news outlets) which can be in one of three possible states -- susceptible, infected, or recovered. Directed edges represent probability of infection (information) transmission from infected to susceptible nodes. Infected nodes become recovered. The model has been recently extensively explored in the fields of social network analysis (to describe spreading of influence in social networks~\cite{Chen2017}, to predict information diffusion probabilities~\cite{Varshney2017}) and socio-physics (to model spreading of emotions~\cite{Xiong2018} and reemergence of information diffusion~\cite{Yang2018}). Here we apply it to model the process of articles diffusion between news outlets. Heuristically, whenever an outlet publishes an original news item (or copies the article from an unobserved source), the competing outlets might decide to publish it as well which in turn might lead to another re-use of the article by theirs neighbors; moreover, if more than one neighbor of a node becomes infected then it is more likely the node will become infected which is a decent representation of a peer pressure between competitors. 

We use the \textit{independent\_cascade} add-on to the \textit{networkx}~\cite{netx} Python library written by Hung-Hsuan Chen. The outline of the algorithm using the SIR models-related terminology is as follows:
\begin{enumerate}
\item All nodes start \textit{susceptible},
\item Change status of one randomly chosen node (source) to \textit{infected},
\item $t=0$,
\item While(number of infected nodes $> 0$):
\begin{enumerate}
\item $t = t + 1$,
\item Each infected node A tries to infect each of its susceptible neighbors B with probability $p_{AB}$,
\item Each node infected in step $t-1$ becomes \textit{recovered}.
\end{enumerate}
\end{enumerate}

The process was simulated on the network extracted from Event Registry data (\textit{real}, see Methods), and two types of synthetic networks -- a random graph (\textit{ER} -- Erd\H{o}s-R\'enyi) and a Barab\'asi-Albert network (\textit{BA}). Sizes of artificial networks were set to be the same as in the giant component of the pruned network ($N=1,037$). The probability of connection between nodes in ER graph was set to be equal to the density of the component ($p=\rho\approx 0.004$, $\langle k \rangle=Np>1$). In the BA graph, we assumed that the starting number of nodes $d_0=d$ and calculated $d = \langle k \rangle/2 \approx 4$. Each edge in generated undirected graphs was changed to two directed edges - one in each of directions. We considered three above-mentioned networks with both homogeneous (\textit{const}; $d_{ij}=0.02$) and heterogeneous edge weights (\textit{shuffled}, drawn from the empirical distribution -- for the three networks; \textit{real}, calculated from the data -- for the recovered network). This sums up to seven variants of a network topology and edge weights -- \textit{real real}, \textit{real shuffled}, \textit{real const}, \textit{ER shuffled}, \textit{ER const}, \textit{BA shuffled}, \textit{BA const}.

\begin{figure}
\centering
    \includegraphics[width=.75\textwidth]{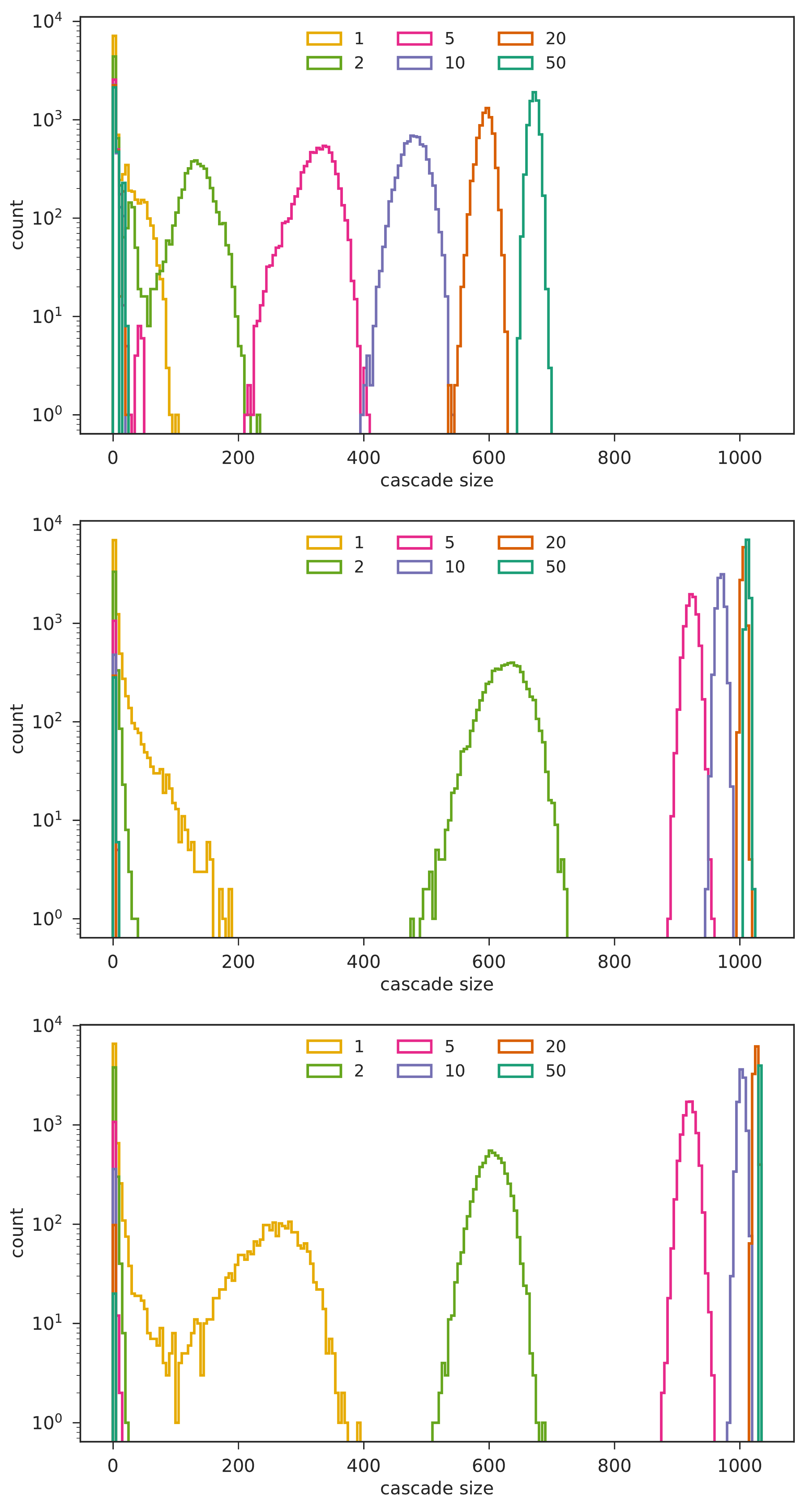}
\caption{Cascade sizes distributions. It is visible that the hype parameter $h$ controls expected size of cascade and chance to exceed viral threshold. Histograms of simulated cascade sizes for different values of multiplicative factor $h$ on (top) the extracted network, (center) a random graph, (bottom) Barabasi-Albert graph. Y-axis is logarithmic. Simulations were performed 10,000 times in each setting.}
\label{fig:hypes_sizes}
\end{figure}

When we assumed that probabilities of link activation are the weights given by Eq. \ref{eq:dij}, \textit{i.e.} $p_{AB}=d_{AB}$, then simulated cascades were rather small (as should be expected for coverage of local and everyday news). To remove this discrepancy we further assumed that the news attractiveness might be captured by a real multiplicative factor $h>0$ (\textit{hype}). The edge weights $d_{AB}$ were multiplied by the selected hype: $p_{AB} = \min(h\,d_{AB},1)$. In Fig.~\ref{fig:hypes_sizes}, there are histograms of cascade sizes for the three networks with shuffled empirical weights with $h\in \{ 1,2,5,10,20,50 \}$. $h=2$ was enough to generate cascades of sizes comparable to half of the network size, and $h=50$ caused all nodes reachable from the source to be infected in every simulation (as expected -- $p_{AB}<0.02$ were discarded). For higher $h$, more cascades exceeded the viral threshold and the expected size of the viral cascade was higher while its variance was lower.

We ran the independent cascade model on each of the selected networks and then performed the fluctuation scaling exponent calculations. We performed the simulations in batches of $10,000$ cascades in a few variants of the hype parameter. Simulation in the first group of variants (\textit{C}) were conducted using the same $h$ value for each cascade (\textit{Cx} meaning $h=x$; the simulations correspond to those presented in Fig.~\ref{fig:hypes_sizes}); in the second group of variants (\textit{P}) the hype parameter was selected at random from the power law distribution with the exponent $\beta$ normed for hypes from 1 to 100 (\textit{Px}: $\beta=-x$); the third group (\textit{TP}) is similar to the second one but the simulations are performed in a sequence from the lowest to the highest values of $h$ mimicking temporal correlations of the hype parameter; the last variant consisted of samples drawn from a uniform distribution in the range of 1--50 (\textit{uni}). The number of hype variants totals to 20 (\textit{C1, C2, C5, P1, P1.5, P2, P2.5, P3, P3.5, P4, P4.5, TP1, TP1.5, TP2, TP2.5, TP3, TP3.5, TP4, TP4.5, uni}).

\begin{figure}
\centering
    \includegraphics[width=\textwidth]{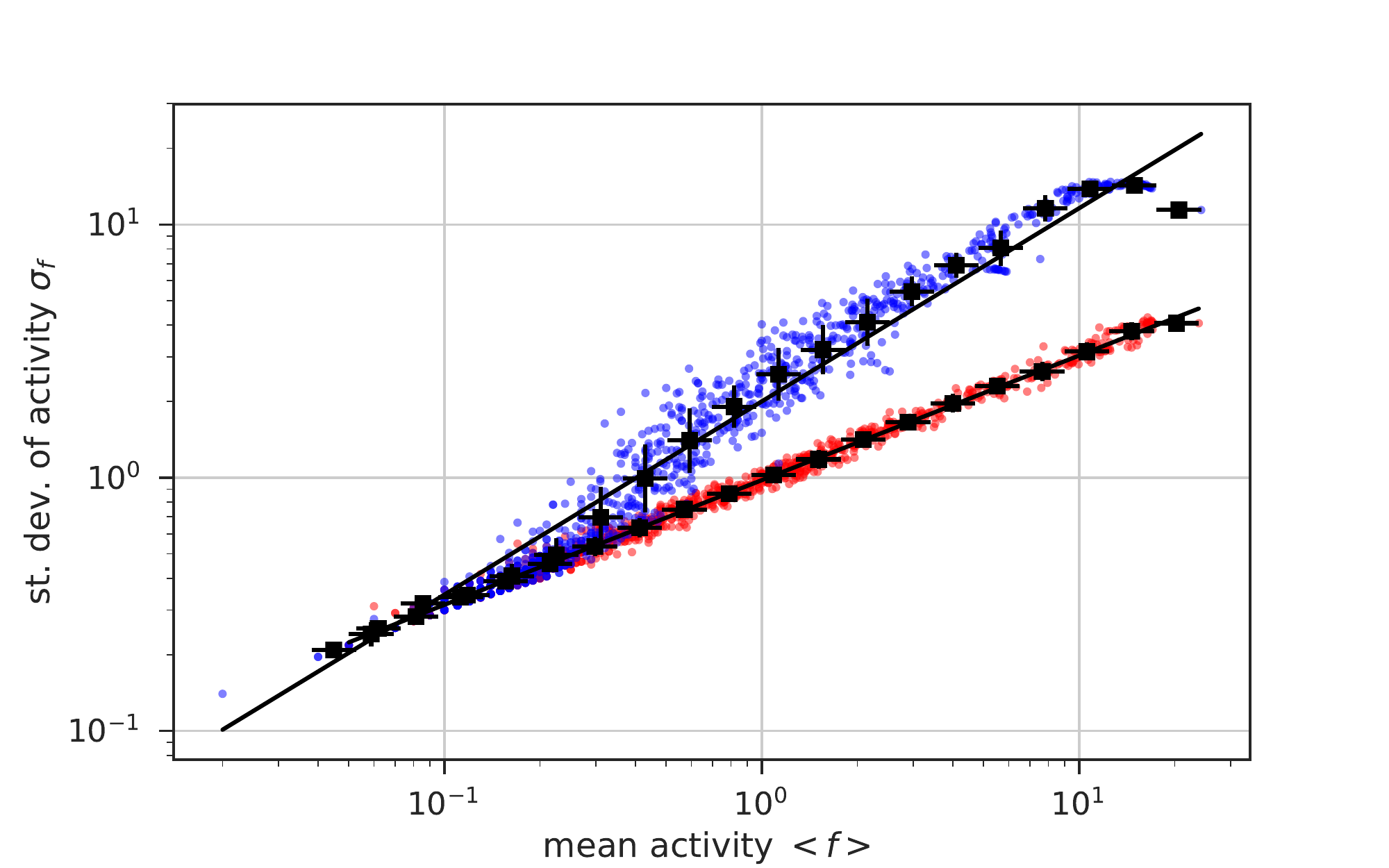}
\caption{Evidence of fluctuation scaling in a model of publishers network. Plots for $\Delta = 100$ and $p(h) \sim h^{-4}$ ($1\leq h \leq 50$) (red points) without temporal hype correlations, (blue points) with temporal hype correlations. The recovered network with the recovered weights was an environment for both simulation batches. Each of simulation batches had $10,000$ realizations. Slopes are (top) $0.492\pm0.002$, (bottom) $0.763\pm0.025$.}
\label{fig:single_model_tfs}
\end{figure}

\begin{figure}
\centering
    \includegraphics[width=.65\textwidth]{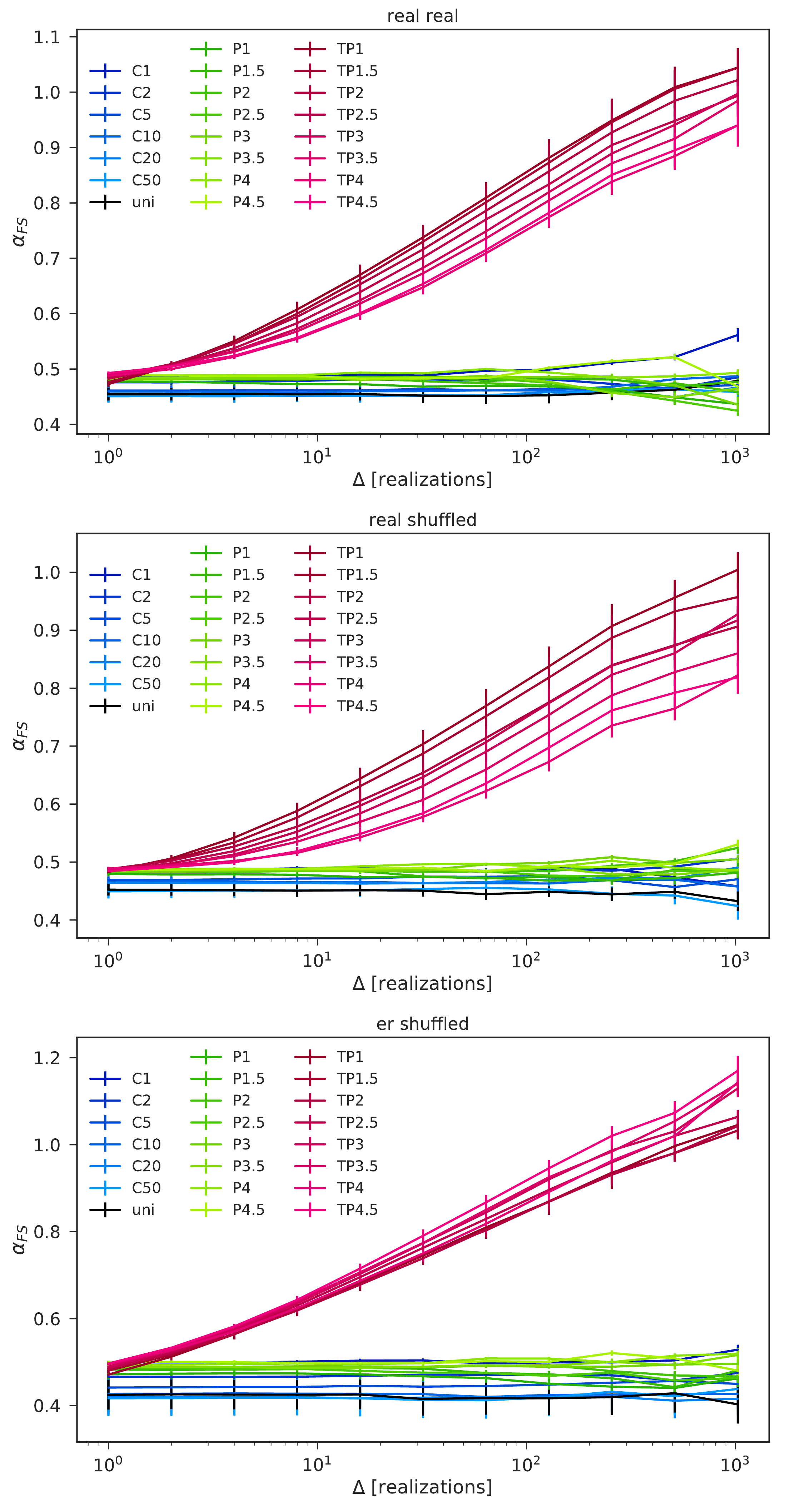}
\caption{Temporal fluctuation scaling in model for different hype distributions ($1\leq h \leq 100$) in the recovered network (top), the recovered network with shuffled edge weights (middle), a random graph with empirical weights (bottom). Only models where edge weights were drawn from the empirical distribution and where simulations with similar $h$ values were grouped (temporal clustering) indicate non-trivial fluctuation scaling exponents. Each color line stands for different hype distributions and orderings (see Section~\ref{sec:model}). Each of simulation batches had $10,000$ realizations.}
\label{fig:model_tfs}
\end{figure}

The temporal fluctuation scaling was observed for all network variants but one (\textit{BA shuffled}). Exemplary plots of the fluctuation scaling observed in \textit{real real} network are shown in Fig.~\ref{fig:single_model_tfs} ($\Delta=100$, top -- \textit{P4}, bottom \textit{TP4}). Moreover, in the group \textit{TP} for \textit{real real}, \textit{real shuffled}, and \textit{ER shuffled} networks, we found that the fluctuation scaling exponent depends on the window size $\Delta$ (Fig.~\ref{fig:model_tfs}). In all other cases, the scaling exponent was very close to $0.5$ for all window sizes (\textit{ER const}, \textit{real const}), or the activity range was insufficient (less than one decade) to meaningfully recover $\alpha$ (\textit{BA const}, \textit{BA shuffled}).

Interestingly, there is barely any difference between results for the recovered network and for the recovered network with shuffled weights; results are also similar for the \textit{ER shuffled} but with a modest activity range (1.5 decade). The character of the dependence is similar to the one obtained for the real data in the previous sections. For networks with homogeneous edge weights, numerical simulations gave $\alpha(\Delta)\approx 0.5$ which suggests that the existing diversity of interaction strengths between publishers is responsible for the observed dependence of the $\alpha$ exponent on the length of the window size $\Delta$ (see Fig.\ref{fig:tfs_dt}).

We conclude the relationship between the fluctuation scaling exponent and the aggregation window size can be received from the independent cascade model run on a heterogeneous network with a slowly changing hype parameter controlling spreading rate.

\section{Discussion}
\label{sec:discussion}
In the paper, we present analyses of a dataset consisting of 22 million articles gathered by EventRegistry.org in 2016 containing at least one of 11 keywords published by over $10,000$ news outlets from around the world.

First, we show long-tailed distributions found in the dataset. Distributions of number of articles published by different outlets can be described using the Weibull distribution; for distributions of number of publishers in different events, and event sizes the log-normal functions were the best fit.

Second, we consider the temporal fluctuation scaling of news outlets' activity around certain keywords. The result put the system on a long list of complex systems following the Taylor's law~\cite{Eisler2008}. The data follows nontrivial fluctuation scaling with three recognized regimes ($\Delta<15\mathrm{ min}$, $15\mathrm{ min}<\Delta<1\mathrm{ day}$, $\Delta>1\mathrm{ day}$). The result suggests that there are different dynamics governing different timescales -- barely any correlations for short timescales, and a varying amount of synchronization for longer timescales. 

Third, we uncover a network of correlations between news outlets content basing on co-occurrences in events and microclusters. The revealed network has interesting features -- e.g. the in-degree distribution is wider than the out-degree distribution, geographical clustering, few strongly correlated groups of sources.

Fourth, we run the independent cascade model on the reverse engineered network and compare it to similar processes run on two synthetic networks (random graph and Barabasi-Albert). Although the independent cascade model leads to the temporal fluctuation scaling for nearly all cases, however to obtain nontrivial exponents observed in the data it is necessary to introduce a multiplicative parameter for a transmission chance (\textit{hype}). Long-tailed event sizes distributions can be obtained using a long-tailed distribution of \textit{hypes} as there is an expected cascade size for a given \textit{hype} for a given network. Moreover, introducing grouping cascades with similar hype yields with an $\alpha(\Delta)$ dependence similar as in the real data. We stress that the uncovered network gives a much better fit to the fluctuation scaling observed in the data as compared to the investigated synthetic networks. 

The above analyses show a few interesting features of the dataset. Statistical inspection suggests the dynamics of news publishing is similar for each outlet depending mostly on a general activity of the outlet on a certain topic. The presence of long-tailed distribution seems to be a universal feature in human online communication channels~\cite{Garas2012, Sienkiewicz2013}, or more generally speaking in complex systems~\cite{sornette2009}. The global news network follows the temporal fluctuation scaling law which is unsurprising as the system consists of spatially/temporally correlated units connected with overlapping communities~\cite{Petri2013}. Collective effects are stronger for longer timescales what corresponds to burstiness of media attention. Basing on our model the specific values of estimated scaling exponents are probably given by the structure of the underlying communication/mimicking network and the distribution of \textit{hype} factor among stories. The proposed \textit{hype} parameter might be interpreted as an external field coupled to observed outlets activity~\cite{fronczak}. The results could be used to meaningfully estimate an impact of a given online story or an influence of a news outlet. 

The presented model is surely not the only way to recover the fluctuation scaling similar to one observed in the data (in general there may be arbitrary many models indicating a given fluctuation scaling) but it shows a role of content attractiveness in information spreading and its fluctuation scaling and provides an interesting interpretation to the uncovered network. The method used to uncover a network of publishers might be an interesting tweak to existing methods of network recovery for cases when the original source or diffusion path is unclear. Our model focuses on propagation of news on a specific topic and does not take into account interests of news outlets. A model considering the whole news flow should definitely include eagerness of a given news outlet to cover given topic (e.g. in a form of news' topic vectors and outlets' interests vectors~\cite{Yu2017}). Moreover, it would be interesting to broaden a range of analyzed time windows but it was impossible for the real data and very time-consuming for the model.
It might be fruitful to consider temporal aspects of links between outlets or even treat the system as a coevolving network~\cite{Toruniewska2017}. Also ego-networks of publishers and community structure of the network might be worth a closer look. An application of recent advancements in cross-lingual text comparisons (e.g.~\cite{rupnik2016}) could lead to uncovering the global content correlation networks.

\appendix
\section{Basic statistical properties of investigated datasets}
\label{app:fits}
We observed that logarithmically-binned density histograms of publisher activities, event sizes, and event coverages are long-tailed. To find the best fit in each case, we considered the following discrete positive ($x\in\{0,1,2,\ldots\}$) distributions provided by the \textit{powerlaw} Python library~\cite{pythonpowerlaw}: power law with exponential cut-off $f(x;\alpha,\kappa)\sim x^{-\alpha}e^{-\kappa x}$, where $\alpha,\kappa>0$;
positive log-normal $f(x;\mu,\sigma)\sim x^{-1} \exp(-(\ln{x} - \mu)^2 / 2\sigma^2)$, where $\mu,\sigma>0$;
Weibull $f(x;\beta,\lambda)\sim (x\lambda)^{\beta-1} 
\exp(-(\lambda x)^\beta)$, where $\beta,\lambda>0$.

Distribution parameters were calculated using corresponding maximum likelihood methods. To determine which of the distributions describes a given histogram most accurately, the fitted functions were compared pairwise using the log-likelihood ratios~\cite{Clauset2009}. For any given pair, a likelihood of the data was calculated under each of competing distributions separately, then the Vuong's log-likelihoods ratio test~\cite{Vuong1989} was performed to determine which distribution was a better fit and whether the result was statistically significant.

\begin{figure}
\centering
\includegraphics[width=\textwidth]{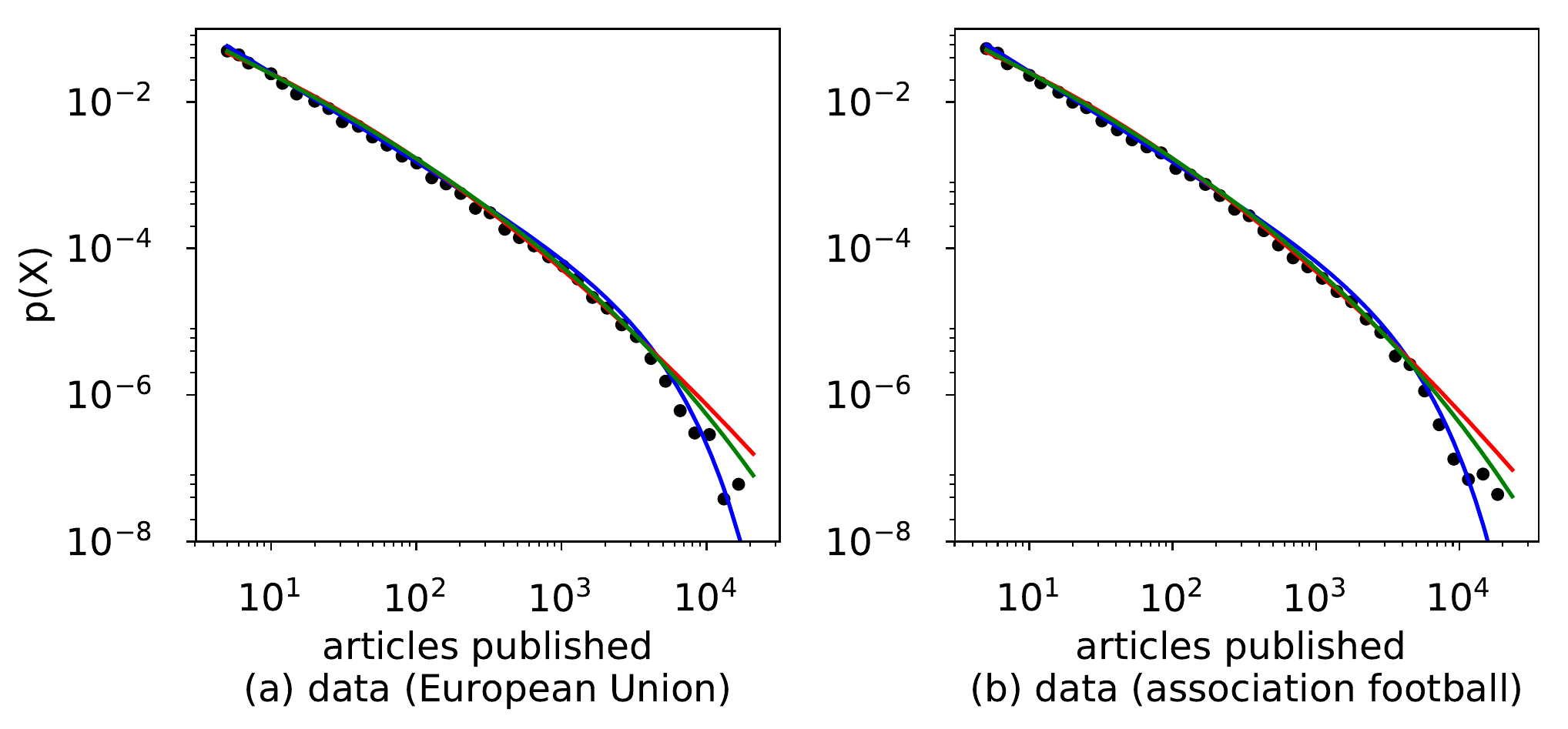}
\caption{Normed histograms of \textbf{publisher activities} for concepts (left) \textit{European Union} and (right) \textit{association football} published by each source. X-axis -- a number of articles with a given concept published by a source, Y-axis -- a normalized count. Data grouped in logarithmic bins, color lines are various types of fitted heavy-tailed distributions (blue -- truncated power law, red -- log-normal, green -- Weibull. The Weibull distribution turned out to be the best fit across majority of analyzed keywords. Fit parameters are in Tab.~\ref{tab:dist_fits}}
\label{fig:pubs_vol_hist}
\end{figure}

For most concepts, the Weibull distribution was the better fit to histograms of publisher activities than the log-normal ($p<0.05$) and the truncated power-law ($p<0.05$) distributions. For the sake of comparability of the results among concepts, parameters for the Weibull distribution ($\beta_A$, $\lambda_A$) were provided for all concepts. Figure~\ref{fig:pubs_vol_hist} shows histograms of publisher activity for keywords \textit{European Union} and \textit{association football}. 

\begin{figure}
\centering
\includegraphics[width=\textwidth]{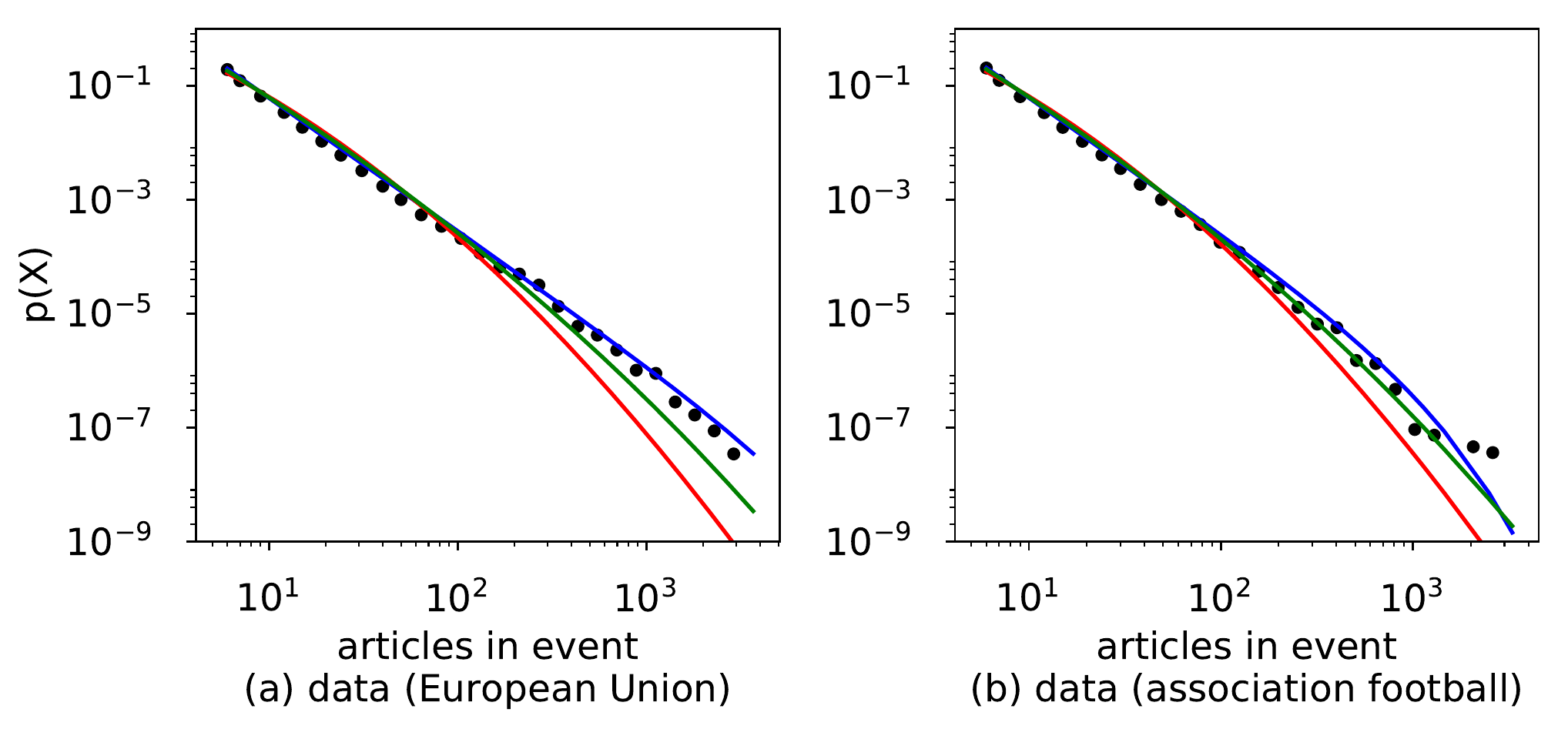}
\caption{Normed histograms of \textbf{sizes of events} containing concepts (left) \textit{European Union} or (right) \textit{association football}. X-axis -- a number of articles with a given concept assigned to an event, Y-axis -- a normalized count. Data grouped in logarithmic bins, color lines are various types of fitted heavy-tailed distributions (blue -- truncated power law, red -- log-normal, green -- Weibull. The log-normal distribution was selected the best fit across majority of analyzed keywords. Fit parameters are in Tab.~\ref{tab:dist_fits}}
\label{fig:events_vol_hist}
\end{figure}

In case of event sizes, the statistical comparisons were inconclusive and we were unable to differentiate between the Weibull, log-normal, and truncated power law functions in all but two cases. The log-normal distribution ($\mu_{EV}$, $\sigma_{EV}$) was chosen to be displayed in an aggregate table (Tab.~\ref{tab:dist_fits}). Figure~\ref{fig:events_vol_hist} presents histograms of sizes of event about \textit{European Union} and \textit{associated football} in terms of article count (event size).

\begin{figure}
\centering
\includegraphics[width=\textwidth]{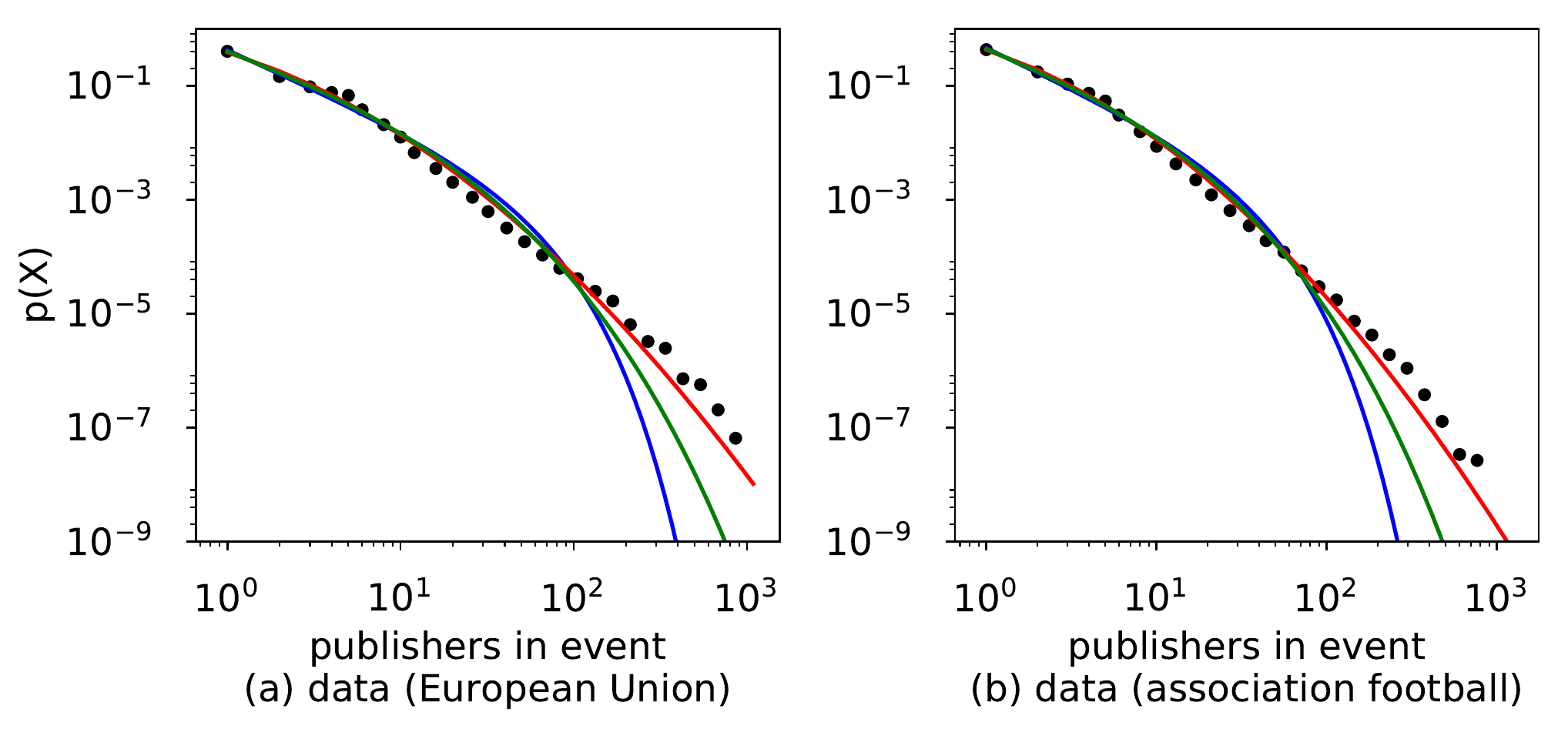}
\caption{Normed histograms of \textbf{coverage of events} containing concepts \textit{European Union} (left) and \textit{association football} (right). X-axis -- a number of sources which published an article assigned to an event, Y-axis -- a normalized count. Data grouped in logarithmic bins, color lines are various types of fitted heavy-tailed distributions (blue -- truncated power law, red -- log-normal, green -- Weibull. The log-normal distribution was selected the best fit across majority of analyzed keywords. Fit parameters are in Tab.~\ref{tab:dist_fits}}
\label{fig:events_coverage_hist}
\end{figure}

For the majority of the analyzed concepts the log-normal distribution was the best fit to the empirical distribution of event coverages (with $p<0.05$ for 7 out of 11 concepts) thus $\mu_{EC}$ and $\sigma_{EC}$ are provided for each concept for a comparison. Histograms in Fig.~\ref{fig:events_coverage_hist} present distributions of sizes of events about \textit{European Union} and \textit{associated football} in terms of involved publishers.

\begin{table}
\centering
\begin{adjustbox}{center}
\begin{tabular}{l|cc|cc|cc}
\toprule
{concept} & $\beta_A$ & $\lambda_A$ & $\mu_{EV}$ & $\sigma_{EV}$ & $\mu_{EC}$ & $\sigma_{EC}$ \\
\midrule
Barack Obama         &    $0.25$ &      $0.35$ &  $1.9\times10^{-8}$ &        $1.70$ &   $3.4\times10^{-8}$ &        $1.66$ \\
Hillary Clinton      &    $0.23$ &      $0.67$ &  $4.1\times10^{-7}$ &        $1.74$ &   $4.7\times10^{-8}$ &        $1.77$ \\
Donald Trump         &    $0.24$ &      $0.33$ &  $1.0\times10^{-8}$ &        $1.72$ &   $0.0636$ &        $1.79$ \\
European Union       &    $0.24$ &      $0.28$ &  $1.4\times10^{-7}$ &        $1.50$ &   $0.2113$ &        $1.49$ \\
United Kingdom       &    $0.28$ &      $0.08$ &  $7.0\times10^{-8}$ &        $1.48$ &   $1.8\times10^{-9}$ &        $1.46$ \\
Germany              &    $0.26$ &      $0.15$ &  $1.0\times10^{-7}$ &        $1.48$ &   $0.0083$ &        $1.46$ \\
France               &    $0.27$ &      $0.11$ &  $2.2\times10^{-8}$ &        $1.50$ &   $0.0762$ &        $1.49$ \\
Argentina            &    $0.21$ &      $1.57$ &  $6.7\times10^{-8}$ &        $1.45$ &   $0.3077$ &        $1.41$ \\
Poland               &    $0.25$ &      $1.00$ &  $5.9\times10^{-7}$ &        $1.45$ &   $1.5\times10^{-7}$ &        $1.38$ \\
democracy            &    $0.27$ &      $0.24$ &  $4.2\times10^{-7}$ &        $1.40$ &   $0.0359$ &        $1.40$ \\
association football &    $0.27$ &      $0.15$ &  $1.8\times10^{-7}$ &        $1.44$ &   $0.1149$ &        $1.37$ \\
\bottomrule
\end{tabular}
\end{adjustbox}
\caption{Fitted parameters of distributions. $\beta_A,\lambda_A$ -- Weibull fit parameters for distribution of publisher activities, $\mu_{EV},\sigma_{EV}$ -- log-normal fit parameters for distribution of event sizes, $\mu_{EC},\sigma_{EC}$ -- log-normal fit parameters for distribution of events coverage (number of unique publishers).}
\label{tab:dist_fits}
\end{table}

\section{Temporal fluctuation scaling}
\label{app:tfs}
The temporal fluctuation scaling has been applied to our data as follows. Let $f_{i,t}$ be a positive variable describing an additive measure of an activity of the object $i$ at time moment $t$.  Examples of such activities can be a number of data packages coming to a router, emails sent by a person, or articles published by an outlet.  Let the total number of elements  in time series of this activity be $T$, i.e. $t=1,2,3, \dots, T$ (further  we will assume that $T$ is the same for all units $i$). Let us further divide the series into $Q$ windows of size $\Delta$, i.e., $Q \Delta =T$. 
The quantity $f_i^{(q,\Delta)}$ stands for a cumulative value of the variable $f_i$ in a  window of size $\Delta$ ($q=1,2,3, \dots,Q$ is the window's label) and $(\sigma_i^\Delta)^2$ is the variance of this cumulative variable in the whole data series. 
Then we have
\begin{equation}
\label{eq:sigma}
(\sigma_i^{(\Delta)})^2= \left\langle[f_i^{(q,\Delta)}]^2\right\rangle- \left\langle[f_i^{(q,\Delta)}]\right\rangle^2
\end{equation}
Here 
\begin{equation}
 \left\langle[f_i^{(q,\Delta)}]\right\rangle = Q^{-1} \sum_{q=1}^{Q}\sum_{t=(q-1)\Delta+1}^{q\Delta}f_{i,t}= \Delta\frac{\sum_{t=1}^{T}f_{i,t}}{T}
\end{equation}
and
\begin{equation}
 \left\langle[f_i^{(q,\Delta)}]^2\right\rangle={1}/{Q}\sum_{q=1}^{Q}\left(\sum_{t=(q-1)\Delta+1}^{q\Delta}f_{i,t}\right)^2
\end{equation} 
It was observed for router activity and email traffic (but also stock markets, river flows, or printing activity)~\cite{Eisler2008} that:
\begin{equation} 
 \sigma_i^{(\Delta)}\propto\left\langle[f_i^{(q,\Delta)}]\right\rangle^{\alpha(\Delta)}. \label{Taylor}
\end{equation}

In practice, data loosely follows the Taylor scaling law thus in order to estimate the exponent $\alpha(\Delta)$ the following procedure was applied:
\begin{enumerate}
\item Calculate mean $\log\langle f\rangle$ for data equally binned by $\log\sigma_i$,
\item Perform least squares fit to the binned data.
\end{enumerate}

The exponent $\alpha$ usually depends on $\Delta$ and few linear regimes can be observed -- the systems observed in~\cite{Eisler2008} followed two, and news outlets activity as described in this study consistently followed three. We used the piecewise linear fit Python library \textit{pwlf} which is the C. Jekel's implementation of the Least-squares Fit of a Continuous Piecewise Linear Function~\cite{pwlf}.

To guarantee sensible statistics for all analyzed units (publishers), we discarded those which had on average less than one article mentioning a given keyword per week in the analyzed period (thus the activity threshold is equal to $52$). Units with mean activity below such a threshold also follow the fluctuation scaling law but with the trivial exponent $\alpha=0.5$; the effect is caused by a relative sparsity of the signals~\cite{meloni2008,pretaylor2016}.

\section{Extracting publishers network}
\label{app:network}
A common approach to uncover the underlying propagation network would be to use information about publication time~\cite{PougetAbadie2015}. Because of the incomplete knowledge about all sources of articles and not fully reliable timestamps, in our case it is hard to determine the original source of a given piece of content~\cite{kivela2012}. We decided to use a co-occurrence fraction counting method known from the field of scientometrics~\cite{Leydesdorff2017}. Thus we calculated similarities between articles in each event to find clusters of highly similar articles (\textit{cascades}) to track which news outlets frequently co-occur in the cascades. Event Registry clustering functionality reduced the required number of pairwise comparisons by a few orders of magnitude. 

To extract cascades from the data, the following procedure was applied to each event from the given period:
\begin{enumerate}
\item download articles,
\item transform each article to a vector of 3-gram occurrences with TF-IDF weighting (trained on a set of $10,000$ randomly selected articles from a week preceding the publication date),
\item calculate a cosine similarity matrix and use it as a distance matrix for a single-linkage hierarchical clustering of the articles,
\item obtain clusters at the threshold value set to $0.25$ (it should guarantee reliable results -- see ~\cite{jch}),
\item save a list of unique publishers in each cluster.
\end{enumerate}
The procedure was implemented using the \textit{scikit-learn} Python library~\cite{scikit-learn}. We decided to use a list of unique publishers because many similar articles from the same source might be caused by the crawler malfunction (e.g. incorrectly obtained article body) or a few updates of the same text. Event Registry system uses filtered stems, concepts, and other metadata; here we used 3-grams to focus on correlations of the content (not only topic) of the articles. 

The publishers network was modeled using all $14,851$ events with articles in English language published between 01-08.05.2017. In the dataset, we found $22,525$ cascades with at least two involved publishers (the total number of cascades was $168,725$).

Let $C$ be a number of observed cascades, $P$ -- total number of publishers contributing.
Following~\cite{Leydesdorff2017}, we construct an occurrence counting matrix $A=\left[a_{pc}\right]$ with $P$ rows and $C$ columns. The matrix $A$ can be also seen as an adjacency matrix of a bipartite graph where the two types of nodes are publishers and cascades (like papers and authors~\cite{Newman2001}, people and social media platforms~\cite{Mitrovi2012}, companies and economic sectors~\cite{Chmiel2007}). Each element of matrix $A$ is defined as: 
\begin{equation}
a_{pc} = \begin{cases}
    1, & \text{if the $p$-th publisher has an article in the $c$-th cascade}.\\
    0, & \text{otherwise}.
  \end{cases}
\end{equation}

The above definition allows us to retrieve a fractional counting cascade co-occurrence $P\times P$ symmetric matrix $U=\left[u_{ij}\right]$ as:
\begin{equation}
u_{ij} = \sum_{c=1}^{C}\frac{a_{ic}a_{jc}}{n^2_c}
\end{equation}
where $n_c$ -- size of the $c$-th cascade. Moreover, the diagonal elements $u_{ii}=N_i$ are fractional occurrence counts for the $i$-th publisher.

We define an asymmetric matrix $D=\left[d_{ij}\right]$ representing a weighted adjacency matrix of the directed publishers network:
\begin{equation}
\label{eq:dij}
d_{ij} = {u_{ij}}/{N_i}
\end{equation}
Trivially, $(\forall_{i,j})(N_i\geq u_{ij})$ as each publisher co-occurred with itself in all cascades it was involved. This means $d_{ij}\leq 1$ and it allows us to use the matrix $D$ as an input for the independent cascade model where $d_{ij}$ will be further used to calculate a probability of an activation of a directed edge from $i$-th to $j$-th publisher assuming that the $i$-th publisher was infected in the previous model step.

The weighted out-degree of the \textit{i}-th node is: 
\begin{equation}
k^{\rightarrow}_i=\sum_{p=1,p\neq i}^{P}d_{ip}
\end{equation}
and it is equal to an expected number of edges activated by the $i$-th node at time $t+1$ if it was infected at time $t$. On the other hand the weighted in-degree of the \textit{i}-th node is: 
\begin{equation}
k^{\leftarrow}_i=\sum_{p=1,p\neq i}^{P}d_{pi}
\end{equation}
and it is equal to an average number of times it would be infected at time $t+1$ if its nearest neighborhood is completely infected at time $t$.

\section*{Acknowledgements}
This research has received funding as {\it RENOIR} Project from the European Union’s Horizon 2020 research and innovation programme under the Marie Sk{\l}odowska-Curie grant agreement No. 691152, by Ministry of Science and Higher Education (Poland), grant Nos. W34/H2020/2016, 329025/PnH/2016, and by National Science Centre, Poland Grant No. 2015/19/B/ST6/02612. J.A.H. was partially supported by the Russian Science Foundation, Agreement \#17-71-30029 with co-financing of Bank Saint Petersburg. This research was also supported in part by PLGrid Infrastructure.


\section*{Data Availability}
The data that support the findings of this study are available from Event Registry but restrictions apply to the availability of these data, which were used under license for the current study, and so are not publicly available. Data are however available from the authors upon reasonable request and with permission of Event Registry.

\section*{References}

\bibliography{jch-tfs}

\end{document}